\documentclass[11pt]{article}
\title{Quantum Orlicz Spaces in Information Geometry}
\author{R.F. Streater\\Dept of Mathematics\\King's College London\\
Strand, WC2R 2LS}
\date{9 June 2004}

\newtheorem{theorem}{\bf Theorem}
\newtheorem{corollary}[theorem]{\bf Corollary}
\newtheorem{definition}[theorem]{\bf Definition}
\begin{document}
\setlength{\oddsidemargin}{0in}
\setlength{\evensidemargin}{0in}
\maketitle
\begin{abstract}
 Let $H_0$ be a selfadjoint operator such that ${\rm Tr}\,e^{-\beta H_{_0}}$ is of trace class
 for some $\beta<1$, and let ${\cal X}_\epsilon$ denote the set of $\epsilon$-bounded forms,
{\em i.e} $\|(H_{_0}+C)^{-1/2-\epsilon}X(H_{_0}+C)^{-1/2+\epsilon}\|<C$ for some $C>0$.
Let ${\cal X}:={\rm Span}\,\cup_{\epsilon\in(0,1/2]}{\cal X}_\epsilon$.
 Let ${\cal M}$ denote the underlying set of the quantum
 information manifold of states of the form
 $\rho_{_X}=e^{-H_0-X-\psi_{_X}}$, $X\in{\cal X}$.
 We show that
if ${\rm Tr}\,e^{-H_{_0}}=1$,
 \begin{enumerate}
 \item the map $\Phi$,
 \[
 \Phi(X)=\frac{1}{2}{\rm Tr}\,
 \left(e^{-H_0+X}+e^{-H_0-X}\right)-1\]
 is a quantum Young function defined on ${\cal X}$
\item The Orlicz space defined by $\Phi$ is the tangent space of ${\cal M}$
 at $\rho_{_0}$; its affine structure is defined by the $(+1)$-connection of Amari
\item The subset of a 'hood of $\rho_{_0}$, consisting of $p$-nearby
states (those $\sigma\in{\cal M}$ obeying $C^{-1}\rho_{_0}^{1+p}\leq
\sigma\leq C\rho_{_0}^{1-p}$ for some $C>1$) admits a flat affine
connection known as the $(-1)$ connection, and the span of this set is part of
the cotangent space of ${\cal M}$
\item These dual structures extend to the completions in the Luxemburg norms.
\end{enumerate}
\end{abstract}

\section{Introduction}
\subsection{The need for an Orlicz topology}
Let ${\cal H}$ be a separable Hilbert space of infinite dimension,
and ${\cal B}({\cal H})$ the $W^*$-algebra of all bounded operators
on ${\cal H}$. The set $\Sigma$ of all normal states can be
furnished with the trace norm. However, the associated metric is
not a good measure of the distance between states. For example, if
$\rho_{_0}\in\Sigma$ has finite entropy: \begin{equation}
S(\rho_{_0}):={\rm
Tr}\,\rho_{_0}\log\rho_{_0}<\infty,\end{equation}any trace-norm
'hood of $\rho_{_0}\in\Sigma$ contains a dense set of states of
infinite entropy. These states cannot be near $\rho_{_0}$ in any
physical sense. Moreover, if $\left\{\rho(t),t\geq 0\right\}$ is
the dynamics of a system, then we expect that
$S(\rho(t)|\rho(0))<\infty$ for all $t\geq 0$, where
\begin{equation}
S(\sigma|\rho):={\rm Tr}\,\rho(\log\rho-\log\sigma)
\end{equation}
is the relative entropy of Umegaki. This, which is related to the
free energy, should be finite for physical states. We need a
stronger topology, such that a 'hood of $\rho_{_0}$ contains only
states $\rho$ for which $S(\rho|\rho_{_0})<\infty$. In this paper
we show that this goal can be achieved by a norm topology, by
developing an analogue of the $L\,\log L$ space: the unwanted
states near a given state $\rho_{_0}$ are outside the space of
finite norm. The norm is a limiting case of the Schatten cross
norms on spaces of compact operators, which can be regarded as
quantum version of Orlicz spaces. Orlicz spaces were first
introduced into information geometry in the classical case by
Pistone and Sempi \cite{Pistone}.

\subsection{The work of Pistone and Sempi}
These authors are statisticians, who developed a theory of best
estimators (of minimum variance) among all locally unbiased
estimators in non-parametric estimation for classical statistical
theory.

Let $({\cal X},\mu)$ be a measure space, and $f$ the density of a
probability measure equivalent to $\mu$. Thus,
\begin{equation}
f(x)>0\hspace{.2in} \mu \mbox{ almost everywhere, and }{\bf
E}_\mu[f]:=\int_{_{\cal X}}f(x)\mu(dx)=1.
\end{equation}
Let $f_{_0}$ be such a density; we seek a useful family of sets
$N$ containing $f_{_0}$, designed to exclude the states of
infinite entropy, and which can be taken to define neighbourhoods of $f_{_0}$.
We then do the same for each point of $N$, and so on, thus
constructing a topological space.

Consider the class of measures of the form
\begin{equation}
f=f_{_0}\exp\{u-\psi_{f_0}(u)\}
\label{perturbation}
\end{equation}
in which $\psi$, called the free energy, is finite, for all states of a one-parameter
exponential family:

\begin{equation}
\psi_{f_{_0}}(\lambda u):=\log{\bf E}_{f_{_0}\mu}\left[e^{-\lambda u}\right]<\infty
\mbox{ for all }\lambda\in[-\epsilon,\epsilon].
\label{cramer1}
\end{equation}
This implies that all moments of $u$ exist in the probability
measure $\nu=\mu f_{_0}$, and the moment generating function is
analytic in a 'hood of $\lambda=0$. The random variables $u$
satisfying (\ref{cramer1}) for some $\epsilon$ are said to lie in the Cram\'{e}r
class, after the statistician Harald Cram\'{e}r.

The Cram\'{e}r class of random variables was shown by Pistone and
Sempi to be a Banach space, and so, to be complete, when furnished
with the Luxemburg norm
\begin{equation}
\|u\|_{_L}:=\inf\left\{r>0:{\bf E}_\mu\left[f_{_0}\cosh
\frac{u}{r}\right]-1\right\}.
\end{equation}
The map
\begin{equation}
u\mapsto\exp\left\{u-\psi_{f_o}(u)\right\}f_{_0}:=f_{_0}(u)
\label{cramermap}
\end{equation}
maps the unit ball in the Cram\'{e}r class into the class of
probability distributions that are absolutely continuous relative
to $\mu$. The identification of $\psi$ as the free energy can be
seen when we write $f_{_0}=\exp\{-h_o\}$, so that
$f=\exp\{-h_o-u-\psi_f(u)\}$; then $h_o$ appears as the `free Hamiltonian',
and $u$ as the perturbing potential, of the Gibbs state $\mu f$. Random
variables $u$ and $v$ differing by a constant give rise to the same distribution.
The map (\ref{cramermap}) becomes bijective if we fix $u$ such that
${\bf E}_\mu[f_{_0}u]=0$, that is, $u$ has zero mean in the measure
$f_{_0}\mu$. Such a $u$ is called a {\em score} in statistics. The corresponding
family of measures $\mu f_{_0}(\lambda u)$, for $\lambda\in[-\epsilon,\epsilon]$,
is called a one-parameter exponential family. Pistone and Sempi define the
neighbourhood $N$ of $f_{_0}$ to consist of all distributions in some exponential family,
as $u$ runs over the Cram\'{e}r class. They add similar 'hoods for each $f\in N$,
and show that the Luxemburg norms are equivalent on overlapping 'hoods. They thus
construct the information manifold ${\cal M}$, which is modelled on the Banach space of
functions of Cram\'{e}r class; this Banach space is identified with the tangent space at
any $f\in{\cal M}$. The manifold is furnished with a Riemannian metric, the Fisher metric,
which at $f\in{\cal M}$ is the second Fr\'{e}chet differential of $\psi_f(u)$.
The Cram\'{e}r class is a special case of an Orlicz space; we now review this.

\subsection{Young functions and Orlicz spaces}
A {\em Young function} is a convex map $\Phi:{\bf R}\rightarrow {\bf R}^+\cup\infty$ such that
\begin{enumerate}
\item $\Phi(x)=\Phi(-x)$
\item $\Phi(0)=0$
\item $\lim_{x\rightarrow\infty}\Phi(x)=+\infty.$
\end{enumerate}
The epigraph of $\Phi$ is the set of points
$\{(x,y):y\geq\Phi(x)\}$;
it is closed and convex. Then $\Phi$ is lower semicintinuous, and $\lambda\mapsto \Phi(\lambda X)$ is continuous
on any open set on which it is finite \cite{Hiriart}.

\vspace{.1in}
\noindent Examples\\
\begin{eqnarray*}
\Phi_1(x)&:=&\cosh x-1\\
\Phi_2(x)&:=&e^{|x|}-|x|-1\\
\Phi_3(x)&:=&(1+|x|)\log(1+|x|)-|x|\\
\Phi^p(x)&:=&|x|^p\mbox{ defined for }1\leq p<\infty.
\end{eqnarray*}

\noindent Let $\Phi$ be a Young function; then its Legendre-Fenchel dual,
\begin{equation}
\Phi^*(y):=\sup\{xy-\Phi(x)\}
\label{legendre}
\end{equation}
is also a Young function. From Legendre theory, we see that $\Phi^{**}=\Phi$.
For example, $\Phi_2=\Phi_3^*$, and $\Phi^p=\Phi^{q*}$ if $p^{-1}+q^{-1}=1$.

\vspace{.1in}
\noindent Equivalence.\\
We say that two Young functions $\Phi$ and $\Psi$ are equivalent if there exists
$0<c<C<\infty$ and $x_0>0$ such that
\begin{equation}
\Phi(cx)\leq\Psi(x)\leq\Phi(Cx)
\end{equation}
for all $x\geq x_0$. We then write $\Phi\equiv\Psi$. The scale of $x$ is not
relevant. Duality is an operation on the equivalence classes:
\begin{equation}
\Phi\equiv\Psi\Longrightarrow \Phi^*\equiv\Psi^*.
\end{equation}
For example, $\Phi_1\equiv\Phi_2.$

\vspace{.1in}\noindent The $\Delta_2$-class.\\
We say that a Young function satisfies the $\Delta_2$-condition iff there exists $\kappa>0$
and $x_0>0$ such that
\begin{equation}
\Phi(2x)\leq\kappa\Phi(x)\mbox{ for all }x\geq x_0.
\end{equation}
For example, $\Phi^p$ and $\Phi_3$ satisfy $\Delta_2$, but not $\Phi_1$ or $\Phi_2$.
\newpage
\noindent The Orlicz class and the Orlicz space\\
Let $(\Omega,{\cal B},\nu)$ be a measure space obeying some mild conditions, and let
$\Phi$ be a Young function. The {\em Orlicz class} defined by $(\Omega,{\cal B},\nu)$
and $\Phi$ is the set $\hat{L}^\Phi(\nu)$ of real-valued
measurable functions $u$ on
$\Omega$ obeying
\begin{equation}
\int_{_\Omega}\Phi(u(x))\nu(dx)<\infty.
\end{equation}
It is a convex space of random variables, and is a vector space iff $\Phi\in\Delta_2$.
The {\em Orlicz space} associated with $\Phi$ and $\nu$ is
\begin{equation}
L^\Phi:=\left\{\rule{0in}{.25in}u: \Omega\rightarrow{\bf R},\mbox{ measurable },\int_{_\Omega}\Phi(\alpha u(x))
\nu(dx)<\infty\mbox{ for some }\alpha>0\right\}
\end{equation}
It is a vector space of random variables, and is the span of the Orlicz class. Up to sets
of measure zero, $L^\Phi$ is a Banach space when furnished with the {\em Orlicz norm}
\begin{equation}
\|u\|_\Phi:=\sup_v\left\{\int|uv|d\nu:v\in L^{\Phi^*},\int\Phi^*(v(x))d\nu\leq 1\right\},
\label{orlicznorm}
\end{equation}
or with the equivalent {\em gauge norm}, also known as a Luxemburg norm, for any $a>0$:
\begin{equation}
\|u\|_{L,a}:=\inf\left\{r>0:\int\Phi(r^{-1}u(x))\nu(dx)\right\}<a\}.
\end{equation}
By {\em the} Luxemburg norm, denoted $\|u\|_{_L}$ we shall mean the case when $a=1$.
Equivalent Young functions give equivalent norms, and $L^\Phi$ is separable
iff $\Phi\in\Delta_2$.

\vspace{.1in}
\noindent Analogue of H\"{o}lder's inequality\\
\noindent We have the inequality
\begin{equation}
\int|uv|\nu(dx)\leq 2\|u\|_{_L}\|v\|_{_L}.
\label{holderorlicz}
\end{equation}
This leads to
\[
L^\Phi\subseteq \left(L^{\Phi^*}\right)^*.\]

\vspace{.1in}
\noindent Examples.
For $\Phi^p(x):=|x|^p$, the Orlicz space is the Lebesgue space $L^p$, and the dual
Orlicz space is $L^q$, where $p^{-1}+q^{-1}=1$. For $\Phi_1$ we get a non-separable
space, sometimes called the Zygmund space when $\Omega={\bf R}$. It is the dual of
$L^{\Phi_3}$, also known as the $L\log L$ space of distributions of
finite differential entropy.

See \cite{Raoren,KR} for classical Orlicz theory.

\vspace{.1in}
\noindent Squeezing in logarithms\\
\noindent When $\nu$ is discrete with countable support, the Orlicz
spaces associated with $\Phi^p$ are the $p$-summable sequences $\ell^p,\hspace{.2in}
1\leq p\leq\infty$. These form a nested family of Banach spaces, with $\ell^1$
the smallest and $\ell^\infty$ the largest. However, this is not the best way to look at
Orlicz spaces. Legendre transforms come into their own in the context of a
manifold, as a transform between the tangent space and the cotangent space
at each point. There is only one manifold, but many coordinatizations. For the
information manifold ${\cal M}$ of Pistone and Sempi, the points of the manifold
are the probability measures $\nu$ equivalent to $\mu$, and these form a positive cone
inside $L^1(\Omega,\mu)$. This cone can be coordinatized by the Radon-Nikodym derivatives
$f=d\nu/d\mu$. The linear structure of $L^1(\Omega,d\mu)$ provides
the tangent space
with an affine structure, which is called the (-1)-affine structure in Amari's notation.
Amari has suggested that we may also use the coordinates
\begin{equation}
\ell_\alpha(f):=f^{(1-\alpha)/2} \hspace{.3in}-1<\alpha<1,
\label{embed}
\end{equation}
known as the Amari embeddings of the manifold
into $L^p$, where $p=2(1-\alpha)^{-1}$, (since $f\in L^1$, we have
$u=f^{(1-\alpha)/2}\in L^p$). Thus, the Orlicz spaces of all the Young functions $|u|^p$ give
the same topology on the manifold, namely, that of $L^1$. So they do not help
in eliminating states of infinite relative entropy.
These coordinates do
provide us with an interesting family of connections,
$\nabla_\alpha:=\partial/\partial\ell_\alpha$, which define the Amari affine structures.

We do better with the formal limit as $p\rightarrow\infty$. In the discrete
case, the relative entropy is the limit as $\alpha\rightarrow 1$ of the Hasegawa-Petz
$\alpha$-entropies
\begin{eqnarray}
S(g|f)&:=&\sum_x f(x)(\log f(x)-\log g(x))\\
&=&\sum_x\lim_{\alpha\rightarrow 1}(1-\alpha)^{-1}\left(f(x)-f(x)^\alpha g(x)^{1-\alpha}\right).
\end{eqnarray}
It turns out that $S_\alpha(f|g)$ is the `divergence' of the Fisher metric along
the $\alpha$-geodesics. The relative entropy $S(g|f)$ arises as the divergence along
the geodesics provided by the embedding
\[
\ell_1(f):=\log f.\]
Thus the affine structure corresponds to the linear structure of the random variables
$u$ where $f=f_0e^u$, as in the theory of Pistone and Sempi. The topology given by the
corresponding Young function $\Phi_3$ is not equivalent to that of $L^1$, but gives
rise to the smaller space $L\log L$, as wanted.

\noindent Is there a quantum analogue to this theory?
\section{The quantum information manifold}
\subsection{The underlying set of the manifold}
Let ${\cal H}$ be a separable Hilbert space, with ${\cal B}({\cal H})$
denoting the algebra of bounded operators on ${\cal H}$, and denote by $\Sigma_+$
the set of faithful normal states on ${\cal B}({\cal H})$.
In \cite{RFS1} it was suggested that the quantum information manifold ${\cal M}$ in
infinite dimensions should consist of $\rho\in\Sigma_+$
with the property that
there exists $\beta_0\in[0,1)$ such that $\rho^\beta$ is of trace class for all
$\beta>\beta_0$. That is, states $\rho\in{\cal M}$ are a bit smoother than general
density operators, in that some fractional power of $\rho$ is also of trace class.
This condition is satisfied by the temperature states of the harmonic oscillator (for which
$\beta_0=0$) and most elementary systems, as well as quantum field theory,
in a box with periodic boundary conditions. Thus, for given $\beta$, the state must
lie in the class ${\cal C}_\beta$ of Schatten, in the unfashionable case $\beta<1$;
this is a complete metrisable space of compact operator furnished with the quasi-norm
$\rho\mapsto\left({\rm Tr}\,\rho^\beta\right)^{1/\beta}$ \cite{Pietsch}. In \cite{RFS1}
we took the underlying set of the quantum infomanifold to be
\begin{equation}
{\cal M}:=\bigcup_{0<\beta<1}{\cal C}_\beta\cap\Sigma_+.
\end{equation}
All these states have finite von Neumann entropy. In limiting the theory to faithful states,
we are imitating the decision of Pistone and Sempi that the probability measures of
the information manifold should be equivalent to the guiding measure $\mu$, rather than say,
merely absolutely continuous; here, the trace
is the quantum analogue of the measure $\mu$. Given a point $\rho_{_0}\in{\cal M}$,
we seek an analogue of the Cram\'{e}r class at $\rho_{_0}$.

\subsection{The quantum Cram\'{e}r class}
Let us write an arbitrary state $\rho_{_0}\in{\cal M}$ as
\begin{equation}
\rho_{_0}=\exp\{-H_0-\psi_{_0}\}.
\end{equation}
This is always possible, since $\rho_{_0}$ is faithful. The choice of
$H_0$ is ambiguous up to a multiple of the identity, since this can
be absorbed into $\psi$, defined by
\[
\psi_{_0}=\log\,{\rm Tr}\,\exp\{-H_0\}=\log\,Z_0.\]
Thus there is no loss in generality by taking $Z_0=1$.

We perturb a given state $\rho_{_0}$ by adding a potential to $H_0$, in analogy with
the classical theory, where the potential is $u$ as in (\ref{perturbation}).  Suppose that $X$
is a quadratic form such that ${\rm Dom}\,X\supseteq{\rm Dom}\,H_0^{1/2}$ and there exist
positive $a,b$ such that
\begin{equation}
|X(\varphi,\varphi)|\leq a\langle H_0^{1/2}\varphi,H_0^{1/2}\rangle+b\|\varphi\|^2
\label{formbounded}
\end{equation}
for all $\varphi\in{\rm Dom}\,H_0^{1/2}$. Then we say that $X$ is form-bounded
relative to $H_0$. The infimum of all $a$ satisfying (\ref{formbounded}) is called the
$H_0$-form bound of $X$; we shall denote the form bound by $\|X\|_{_K}$, in honour of T. Kato.
It is a semi-norm
on the linear set of forms bounded relative to $H_0$. It is well known that if $\|X\|_{_K}<1$,
then $H_0+X$ defines a semibounded self-adjoint operator. More, if $\|X\|_{_K}$ is small
enough, less that $a(\beta_{_0})$ say, depending on $\beta_{_0}$, then \cite{RFS1}, we have
\begin{equation}
\rho_{_X}:=e^{-H_0-X-\psi_{_X}}\in{\cal M}.
\label{rhoX}
\end{equation}
To prove that $\rho_X$ is of trace class, write $-H_0-X=-\beta H_0-(1-\beta)H_0-X$; then
 by the Golden-Thompson inequality, taking $\beta_{_0}<\beta<1$,
\begin{eqnarray*}
{\rm Tr}\,\rho_{_X}&\leq& {\rm Tr}\,e^{-\beta H_0}e^{-(1-\beta)H_0-X}\\
&\leq&\|\rho_{_0}^\beta\|_1\|e^{-(1-\beta)H_0-X}\|_\infty\\
&<& \infty.
\end{eqnarray*}
More is true \cite{RFS1}; $\rho_{_X}$ inherits the properties of $\rho_{_0}$ with a new
$\beta_{_0}$ nearer 1, and lies in ${\cal M}$.
In \cite{MRGRFS1}, we added a further condition on the quadratic form, called
$\epsilon$-boundedness:
\begin{definition}
{\em For any $\epsilon\in(0,1/2]$ we say that a quadratic form $X$ is
{\em $\epsilon$-bounded} by $H_{_0}$ if there exists a constant $C$ such that
\[ (H_{_0}+C)^{-1/2-\epsilon}X(H_{_0}+C)^{-1/2+\epsilon}\leq CI.\]}
\label{epsilon}
\end{definition}
The set of states satisfying (\ref{epsilon}) is obviously $(+1)$-convex;
that is, if $X_1$ and $X_2$ satisfy (\ref{epsilon}), then so does
$\lambda X_1+(1-\lambda)X_2$. We showed \cite{MRGRFS1} that the free energy is an
analytic function of the perturbation parameter in a small 'hood of zero. This, then,
is an analogue of the Cram\'{e}r condition. Here, we use this condition to
specify the tangent space of ${\cal M}$ at $\rho_{_0}$.  For the cotangent space, we replace
(\ref{epsilon}) by a (possibly) different set, of states $\rho_{_X}$ that are $p$-nearby $\rho_{_0}$,
defined and used in \cite{RFS3}: for some $C>1$,
and $p\in(0,1)$,
\begin{equation}
C^{-1}\rho_{_0}^{1-p}\leq \rho_{_X}\leq C\rho_{_0}^{1-p}.
\label{p}
\end{equation}
The set of states
$\rho_{_X}$ satisfying (\ref{p}) is obviously (-1)-convex:
if $\rho_1=\rho_{_{X_1}}$ and $\rho_2=\rho_{_{X_2}}$ are both $p$-nearby $\rho_0$, then so
is $\lambda\rho_1+(1-\lambda)\rho_2$. It is not known whether it
is $(+1)$-convex unless $p=0$.
It is not hard to show that (\ref{p}) implies that
for small enough $p$,
$\rho_{_X}\in{\cal M}$ \cite{RFS3}. It is easy to show that the intersection
of the sets
\[
\left[\bigcup_{\epsilon>0}\left\{\rho_{_X}: X \mbox{ is }\epsilon \mbox{ bounded}\right\}\right]
\bigcap\left[\bigcup_{p\in(0,1)}\left\{\mbox{states } p\mbox{-nearby }\rho_{_0}\right\}\right]\]
contains the set of states with finite Araki norm \cite{RFS3}; this set carries both
affine structures.

Our strategy in furnishing ${\cal M}$ with a topology
is a quantum version of \cite{Pistone}. We parametrise states near $\rho_{_0}$ by
the potential $X$, and can adjust $X$ and
$\psi_{_X}$ so that the generalised mean $\rho_{_0}\cdot X$ of $X$ in the state $\rho_{_0}$,
proved to be finite in \cite{RFS1},
is zero:
\begin{equation}
\rho\cdot X:={\rm Tr}\,\int_0^1\rho^tX\rho^{1-t}dt=0.
\end{equation}
These are the quantum {\em scores}. The $(+1)$ affine structure on forms
satisfying (\ref{epsilon}) gives, by transfer of structure, an affine structure
to the corresponding part of ${\cal M}$. Thus we
get a piece $N$ of a flat manifold modelled on a vector space. When furnished with
the $\epsilon$-norms, with any point of $N$ replacing $\rho_{_0}$, the norms on overlapping
'hoods are equivalent. We thus get a Banach manifold.
It has the interesting property
that there are no global linear coordinates, even though the coordinate patches are
linear with linear transition functions. To see this, consider perturbations of the form
$X=(k-1)H_0$, which is $H_0$-small enough if $k$ is close to 1. We cannot use the
perturbation if $k=0$, as then $X=-H_0$, and $\exp\{-(H_0+X)\}=I$, which is not of trace class.
Roughly speaking, the manifold is a convex cone
pointing in the general direction of $H_0$. This suggests that the correct Orlicz
space must fail to satisfy the (technical) $\Delta_2$-condition. The Orlicz class
at $\rho_{_0}$,
which is always convex but might not itself be linear, should allow only perturbations
$X$ with sufficiently small norm. Then the Orlicz space, the linear span of the Orlicz class,
parametrises the tangent space of ${\cal M}$ at $\rho_0$ but the scores will not
provide a valid parametrization of the whole manifold.
 Our suggested Young function, below, shows these features.

\subsection{The category of partially ordered Riemannian manifolds}
Amari has posed the question \cite{Amari}, what properties, extra to
being a Riemannian manifold, characterise information manifolds? Obviously, such a
manifold must possess the Amari family of affine connections, $\{\nabla_\alpha\}$, with
$\nabla_\alpha$ dual to $\nabla_{-\alpha}$ relative to the metric. One could ask the
same question for quantum information manifolds. These affine connections are
associated with the embeddings (\ref{embed}), which can be extended to weights
(non-normalised probabilities) and have quantum versions
\begin{equation}
\ell_\alpha(\rho):=\rho^{2/(1-\alpha)}, \hspace{.3in}-1<\alpha<1.
\end{equation}
The quantum versions of the limit cases, $\alpha\rightarrow\pm 1$, are
\begin{equation}
\ell_+(\rho):=\log\rho\mbox { and }\ell_-(\rho):=\rho.
\end{equation}
It is a fact that all these maps are {\em operator monotone}; they preserve
the partial order between operators. We say $A>B$ if $A-B$ is a positive (semidefinite)
operator. Let us say that a coordinate system $\rho\mapsto \ell(\rho)$ for the set
of weights is a monotone coordinate system if this partial ordering is preserved.
An allowed coordinate system for the quantum case must be monotone,
and a morphism between two information manifolds must involve
monotone maps. This differs from Chentsov's definition of morphism;
it allows non-linear changes of coordinates, which transform one monotone metric to another
\cite{Les}. This suggests the following definition:
\begin{definition}
{\em Let ${\cal M}_1$ and ${\cal M}_2$ be Riemannian manifolds with partial orders $\leq _1$
and $\leq _2$. A map $T:{\cal M}_1\rightarrow{\cal M}_2$ is called a {\em morphism of
partially ordered Riemannian manifolds} if $T$ is a morphism of Riemannian manifolds
and maps any two comparable points of ${\cal M}_1$ into comparable points of
${\cal M}_2$, and the order is preserved.}
\end{definition}
This defines the category of partially ordered Riemannian manifolds. For example,
in finite dimensions consider the set of faithful weights in $M^n$ furnished with the
{\em BKM} metric, $g_{_B}$. Then an operator
monotone bijection on this set transforms $g_{_B}$ to another monotone metric, $g$.
According to \cite{Les}, by this means we can get any of the monotone metrics as
classified by Petz \cite{Petz}. Thus all the models are isomorphic when regarded
as partially ordered Riemannian manifolds.

\subsection{The quantum Young function}

In some recent work \cite{Zeg} on quantum Orlicz spaces, use was made of classical
Young functions. Thus, if $X$ is a self-adjoint operator, and $\Phi$ is a Young function,
one can take as the quantum Young function the map $A\mapsto {\rm Tr}\,\Phi(|\tilde{A}|)$,
where $\tilde{A}$ is the rearranged operator. For the cases $-1<\alpha<1$, this gives us
back the trace-norm topology, as explained in the classical case above, when we put $A=\rho$.
In the limit case $\ell_+$, we encounter $\cosh X$, which does not make sense for
forms, and also does not correspond the perturbation by a potential. In the classical case,
$f_0e^{-u}=e^{-h_o-u}$, but in the quantum case, $\rho_{_0}e^{-X}$ is not hermitian,
even if $X$ is a bounded operator, unless $[H_0,X]=0$. The Young function
$\Phi_1=\cosh x-1$ gets multiplied by $f_0$ in the classical theory ({\em c.f.} (\ref{cramer1})),
but the quantum analogue of this would be ${\rm Tr}\,(\rho_{_0}(\cosh|X|-1))$ which
is not positive. We therefore take a different choice of ordering for the
non-commuting variables, and suggest \cite{RFS1} that the quantum Young function at
$\rho_{_0}$ should be
\begin{equation}
\Phi(X)=\frac{1}{2}{\rm Tr}\,\left(e^{-H_0-\psi_o-X}+e^{-H_0-\psi_o+X}\right)-1.
\label{quantumphi}
\end{equation}
If $H_0$ commutes with $X\in{\cal B}({\cal H})$, this reduces to
${\rm Tr}\,\rho_{_0}\Phi_1(X)$.
Since this already includes the factor $\rho_{_0}$, we must omit this factor
in the analogues of (\ref{cramer1}) and the rest.
For $p$-nearby states, we can take the analysis of \cite{RFS1} further:

\begin{theorem}{\em Suppose that $\rho_{_X}$ is $p$-nearby $\rho_{_0}$, for some $p<1-\beta_{_X}$. Then
{\em BKM} regularised metric
\begin{equation}
\langle X,X\rangle_{_B}:=\frac{1}{2}\int_0^1 d\alpha {\rm Tr}\,\left\{\rho_{_0}^{\alpha/2}
X\rho_{_0}^{1-\alpha}X\rho_{_0}^{\alpha/2}\right\}
\end{equation}
is well-defined.}
\label{BKM}
\end{theorem}
PROOF: Since $X=H_{_X}-H_0$, it is enough to consider
the case where each $X$ is replaced by $H_X$, as the remaining terms
involve $H_0$ and are easily bounded.
We suppose that $\rho_{_0}\leq C\rho_{_X}^{1-p}$; since $x\mapsto x^\alpha$
is operator monotone for $0<\alpha<1$, we see that
\[
\rho_{_X}^{\frac{-\alpha(1-p)}{2}}\rho_{_0}^{\alpha}\rho_{_X}^{\frac{-\alpha(1-p)}{2}}\]
is a bounded operator; the same goes if $\alpha$ is replaced by $1-\alpha$. We
write the integrand as the product
\begin{eqnarray*}
\rho_{_0}^\alpha H_{_X}\rho_{_0}^{1-\alpha}&=&\rho_{_X}^{\alpha(1-p)/2}\left(\rho_{_X}
^{-\alpha(1-p)/2}\rho_{_0}^\alpha\rho_{_X}^{-\alpha(1-p)/2}\right)\rho_{_X}^{\alpha(1-p)/2}X\\
&&\rho_{_X}^{(1-p)(1-\alpha)/2}\left(\rho_{_X}^{-(1-p)(1-\alpha)/2}\rho_{_0}^{1-\alpha}\rho_{_X}
^{-(1-p)(1-\alpha)/2}\right)\rho_{_X}^{-(1-p)(1-\alpha)/2}H_X
\end{eqnarray*}
of which the trace (by H\"{o}lder's inequality for traces) is bounded by
\[
C^{\alpha}\left\|H_{_X}\rho_{_X}^{(1-p)/2}\right\|_2\,C^{1-\alpha}\left\|H_{_X}
\rho_{_X}^{(1-p)/2}\right\|_2.\]
The Hilbert-Schmidt norm is finite:
\[\rho_{_X}^{(1-p)/2}H_{_X}=\rho_{_X}^{(1-p-\delta)/2}\left(\rho_{_X}^{\delta/2}H_{_X}\right)\]
for any small $\delta>0$ is the product of a Hilbert-Schmidt operator and a bounded operator
with norms independent of $\alpha$; thus the integral is finite.\hspace{\fill}{\em QED}
\begin{corollary}
{\em The usual proof of the Bogoliubov-Peierls inequality holds, to arrive at the inequality}
\[\log{\rm Tr}\,e^{-H_{_0}+X}\geq \log{\rm Tr}\,e^{-H_{_0}}+\rho_{_0}\cdot X\]
\end{corollary}

\begin{definition}
{\em Let us say that a map $\Phi$, from a linear subspace ${\cal X}$ of the space of
$H_{_0}$-bounded quadratic forms,
to ${\bf R}^+\cup\{\infty\}$ is a quantum Young function for ${\cal X}$ if
\begin{enumerate}
\item $\Phi(X)$ is finite for all forms $X$ with sufficiently small Kato bound
\item $X\mapsto\Phi(X)$ is convex
\item $\Phi(-X)=\Phi(X)$ for all $X\in{\cal X}$
\item $\Phi(0)=0$, and if $X\neq 0$, $\Phi(X)>0$, including $\infty$ as a possible value.
\end{enumerate}}
\end{definition}
\begin{theorem}
{\em For each $\rho\in{\cal M}$, the map $\Phi$ of (\ref{quantumphi}) is a
quantum Young function.}
\end{theorem}
{\bf PROOF} Lemma (4) of \cite{RFS1} gives the proof of (1).\\
For (2), it is known \cite{Lieb} that for self-adjoint $A$, the map
$A\mapsto{\rm Tr}\,e^{A}$ is convex, so that
\[
{\rm Tr}\,e^{\lambda A+(1-\lambda)B}\leq \lambda{\rm Tr}\,e^A+(1-\lambda){\rm Tr}\,e^B.\]
Put $A=-H_0-X$ and $B=-H_0-Y$, where $X$ and $Y$ are sufficiently $H_0$-small forms. Then
$\lambda X+(1-\lambda)Y$is also a sufficiently small form, and
\[
{\rm Tr}\,e^{-H_0-\lambda X-(1-\lambda)Y}={\rm Tr}\,e^{\lambda A+(1-\lambda)B}\leq
\lambda{\rm Tr}\,e^{-H_0-X}+(1-\lambda){\rm Tr}\,e^{-H_0-Y}.\]
Then $\Phi$, being the sum of two convex functions, is convex.\\
Items (3) and (4) are obvious. {\em QED}

\subsection{The Luxemburg norm}
We now specialize to the Young function of interest, associated
with a point $\rho_0\in{\cal M}$. Thus,
$\rho_0=\exp\{-H_0-\psi_0\}$, and $e^{-\beta H_0}$ is of trace
class for some $\beta<1$. Let $Q_0$ be the quadratic form
\begin{equation}
Q_0(\phi):=\|H_0^{1/2}\phi\|^2, \end{equation} and let $X$ be a
$Q_0$-bounded quadratic form.
If $\|X\|_{_K}>1$, then $\Phi(X)$ is put equal to $\infty$, since
either $H_0+X$ or $H_0-X$ is not bounded below. It might be that
even when $\|X\|_{_K}<1$, $\Phi(X)$ is still $\infty$, because
although $H_\pm:=H_0\pm X$ are both self-adjoint and bounded
below, $e^{-H_\pm}$ might not be not of trace class. Let us denote by
$\|X\|_{_k}$ the infimum of the $Q_0$-bounds of $X$ such that
$e^{-H_\pm}\notin{\cal C}_1$, or $\infty$ if $X$ is $Q_{_0}$-tiny. We showed in \cite{RFS1} that
$\|X\|_{_k}>0$. Then we
can define a lower semi-continuous Young function on the
one-dimensional set of forms $\{\lambda X:\lambda\in{\bf R}\}$ by
$\Phi(\lambda X)$ for small enough $\lambda$, and by
\begin{equation}
\Phi(X):=\left\{\begin{array}{cc} \lim_{\lambda\rightarrow
1}\Phi(\lambda X)&\mbox{ \rm if }\|X\|_{_K}=\|X\|_k\\
\infty&\mbox{ \rm if }\|X\|_{_K}>\|X\|_k\end{array}\right.
\label{qyoung2}
\end{equation}

\begin{theorem}
{\em With $\Phi$ given by (\ref{qyoung2}), we
have
\begin{description}
\item[(i)] \[
\|X\|_{_L\,_a}:=\inf_r\left\{r:\Phi\left(\frac{X}{r}\right)<a\right\}\]
defines a norm on ${\rm Span}_{\bf R}{\cal X}$. \item[(ii)]All
these norms, for various $a>0$, are equivalent.
\end{description}}
\label{qnorm}
\end{theorem}
PROOF \\
(i) Obviously, $\|\;\cdot\;\|_{_L\,_a}\geq 0$, and for $\lambda\neq
0$,
\begin{eqnarray*}
\|\lambda X\|_{_L\,_a}&=&\inf\left\{r>0:\Phi\left(\frac{\lambda
X}{r}\right)<a\right\}\\
&=&\inf\left\{|\lambda|s>0:\Phi\left(\frac{X}{s}\right)<a\right\}\\
&=&|\lambda|\,\|X\|_{_L\,_a}.
\end{eqnarray*}
Also, if $X=0$,
\[ \|X\|_{_L\,_a}=\inf\{r>0:\Phi(0)<a\}=\inf\{r>0\}=0.\]
Conversely, if $X$ is such that $\|X\|_{L,a}=0$, then there must
exist a sequence $r_n\rightarrow 0$ such that
\begin{equation}
\Phi\left(\frac{X}{r_n}\right)<a. \label{help}
\end{equation}
 But by assumption, if $X\neq 0$,
$\Phi(sX)>0$ for some $s>0$; convexity then shows that
$\Phi(sX)\rightarrow \infty$ at least as fast as linear in $s$,
contradicting (\ref{help}); this shows that $X=0$.

Finally, for the triangle inequality, put $r=s+t$, $\lambda=s/r$,
$1-\lambda=t/r$. Then the set
\[
A:=A(a):=\left\{r:\Phi\left(\frac{X+Y}{r}\right)<a\right\}\]
contains the set
\[
A_0=\left\{s+t:\lambda\Phi\left(\frac{X}{s}\right)+(1-\lambda)\Phi\left(\frac{Y}{t}\right)
<a\right\}.\] For suppose that $r=s+t\in A_0$. Then
\begin{eqnarray*}
\Phi\left(\frac{X+Y}{r}\right)&=&\Phi\left(\lambda\frac{X}{s}+(1-\lambda)\frac{Y}{t}\right)\\
&\leq&\lambda\Phi\left(\frac{X}{s}\right)+(1-\lambda)\Phi\left(\frac{Y}{t}\right)\\
&<&a,
\end{eqnarray*}
showing that $r\in A$, and so $A_0\subseteq A$. The set $A_0$
contains the set
\[
A_{00}:=\left\{s+t:\Phi\left(\frac{X}{s}\right)<a\mbox{ and
}\Phi\left(\frac{Y}{t}\right)<a\right\}.\] For suppose $r\in
A_{00}$. Then there exist $s,t$ such that $s+t=r$ and
$\Phi(X/s)<a$ and $\Phi(Y/t)<a$. Then there exists $s+t$ such that
\[
\lambda\Phi\left(\frac{X}{s}\right)+(1-\lambda)\Phi\left(\frac{Y}{t}\right)\leq
(\lambda+(1-\lambda))a=a,\] so $r\in A_0$. This shows that
$A_{00}\subseteq A_0\subseteq A$. Since the infimum of a larger
set of real numbers is not greater than the infimum of a smaller
set, we have
\begin{eqnarray*}
\|X+Y\|_{{_L}\,_a}&=&\inf A\leq \inf A_{00}\\
&=&\inf\left\{s+t:\Phi\left(\frac{X}{s}\right)<a\mbox{ and
}\Phi\left(\frac{Y}{t}\right)<a\right\}\\
&=&\inf\left\{s:\Phi\left(\frac{X}{s}\right)<a\right\}+\inf\left\{
t:\Phi\left(\frac{Y}{t}\right)<a\right\}\\
&=&\|X\|_{{_L}\,_a}+\|Y\|_{{_L}\,_a}.
\end{eqnarray*}
This proves (i).

\vspace{.1in}\noindent(ii) We may assume that $a>b$; then
\[
\|X\|_{_L\,_a}\leq \|X\|_{_L\,_b}\] so $\|X\|_{_L\,_b}$ is
the stronger norm. It remains to show that $\|X\|_{_L\,_b}$ is
also weaker. If $X$ is $Q_0$ tiny, when the Kato seminorm
$\|X\|_{_K}$ vanishes, then $\Phi(\lambda X)$ is finite and
continuous, increasing in $\lambda$ to infinity (by convexity). It
therefore passes $a$ and $b$ at points $\lambda=a^\prime$ and
$b^\prime$, where
\[
a^\prime=\|X\|_{_L\,_a}^{-1}\hspace{.6in}{\rm
and}\hspace{.6in}b^\prime=\|X\|_{_L\,_b}^{-1}\] respectively.
From convexity,
\[
\Phi(b^\prime X)=\Phi\left(\frac{b^\prime}
{a^\prime}X+\left(1-\frac{b^\prime}
{a^\prime}\right)0\right)\leq\frac{b^\prime}{a^\prime}\Phi(a^\prime
X).\] Thus
\[
b\leq\frac{\|X\|_{_L\,_a}}{\|X\|_{_L\,_b}}.a\] giving
\begin{equation}
\|X\|_{_L\,_b}\leq\frac{a}{b}\|X\|_{_L\,_a}, \label{weaker}
\label{Orliczab}
\end{equation}
showing equivalence in this case. This set-up, in which the
infimum of the sets $A(a)$ and $A(b)$ are both achieved in
$0<r<\|X\|_k^{-1}$, could also arise if $\|X\|_{_K}>0$, and
leads to the same conclusion.

Now suppose that $\|X\|_{_K}=1$, and consider $\Phi(\lambda X)$ as a function
 of $\lambda$. It is
possible that $b$ is not reached by any $\Phi(\lambda X)$ before
$\lambda=\|X\|_{_k}$, in which case
$\|X\|_{_L\,_b}=\|X\|_{_k}^{-1}$. In that case $a$ is also not
reached by $\Phi(\lambda X)$ before it becomes infinite, and
$\|X\|_{_L\,_a}=\|X\|_{_k}^{-1}$ too, and the norms are equal,
and so equivalent.

The only remaining possibility is that $\Phi(b^\prime X)=b$ for
some $b^\prime<\|X\|_{_k}$, giving $\|X\|_{_L\,_b}=1/b^\prime$,
while $a$ is not reached by $\Phi(\lambda X)$ before
$\lambda=a^{\prime\prime}:=\|X\|_{_k}$, so that
$\|X\|_{_L\,_a}=\|X\|_{_k}^{-1}=1/a^{\prime\prime}$. Then by
convexity,
\begin{eqnarray*}
b&=&\Phi(b^\prime
X)=\Phi\left(\frac{b^\prime}{a^{\prime\prime}}a^{\prime\prime}
X+\left(1-\frac{b^\prime}{a^{\prime\prime}}\right).0\right)\\
&\leq&\frac{b^\prime}{a^{\prime\prime}}\Phi(a^{\prime\prime}X)=
\frac{b^\prime}{a^{\prime\prime}}a.
\end{eqnarray*}
This can be rearranged to give (\ref{weaker}), which completes the
proof of (ii).\hspace{\fill}{\em QED}

In view of (ii), we can take $a=1$, and use the notation
$\|\;\cdot\;\|_{_L}$ for the Luxemburg norm
$\|\;\cdot\;\|_{_L\,_1}$. It is clear that $\| X\|_{_L}\geq
\|X\|_k$: by our convention, $\Phi(X/r)$ is infinite for
$r<\|X\|_k$. This convention is inevitable; for, if both
$\exp\{-H_0\pm X\}$ are of trace class, there exists $C$ such that
${\rm Tr}\,\exp\{-H_0\pm X\}\leq C$. But the state is a positive
operator, so its trace is its trace norm, which is larger that its
operator norm. Hence
\[
0\leq\exp\{-H_0\pm X\}\leq CI.\] Taking logs (an operator monotone
operation, also valid for forms) gives
\[
\pm X\leq H_0+\log C.\] Thus $X$ must be $Q_0$-bounded with bound
$\leq 1$: no larger $\|X\|_{_K}$ can give finite $\Phi$.

It is likely that in our situation, $\Phi(X/r)$ goes smoothly to
infinity as $r\rightarrow 0$, passing through all positive values,
and diverging to infinity at $r=\|X\|_k^{-1}$. If this were
known to be true, then the proof of (ii) would be the same as the
easy case when $X$ is $Q_0$-tiny.

\subsection{Duality}
In \cite{Gib}, the authors associate with a Banach manifold
${\cal M}$ a whole bundle of tangent spaces, coming from the
various Amari embeddings, $\rho\mapsto
\ell_\alpha(\rho)=\rho^{1/p}$. This elegant point of view actually
contains the fact that there is only one tangent space and one
cotangent space, each of which is furnished with a family of
affine connections.

We adopt a more concrete version, mainly because we do not yet
know whether our space is complete, uniformly convex etc. as
required by \cite{Gib}. Let $\rho_{_0}\in{\cal M}$. The set of
states
\[{\cal X}:=\left\{\rho_{_X}:X\mbox{ is $H_{_0}$-$\epsilon$-bounded}\right\}\]  can
be furnished with $(+1)$-affine structure and with the Luxemburg
norm. This space might not be complete. We parametrise the space
by the {\em scores, $X$}. The topological dual ${\cal X}^d$ of the
completed space of scores will contain density operators with
finite entropy, and possibly unwanted non-normal states and weights. We take the
subset ${\cal X}^*\subset{\cal X}^d$ being the $(-1)$-linear span of
density operators obeying (\ref{p}) for some $p$, which, as
remarked, carries the $(-1)$-affine connection. The  pair
${\cal X},{\cal X}^*$ is generated by the Amari embeddings of the
set of states near $\rho_{_0}$:
\[
\rho\mapsto \ell_+(\rho):=\log\rho\mbox{ and its dual }\rho\mapsto
\ell_-(\rho):=\rho\] and their associated affine connections,
$(+1)$ and $(-1)$. We may then write
\[S(\rho_{_0}|\rho_{_X})+S(\rho_{_X}|\rho_{_0})={\rm
Tr}\,\left[\left(\log\rho_{_X}-\log\rho_{_0}\right)\left(\rho_{_Y}-\rho_{_0}\right)\right].\]
We take the limit so that the differences define tangent vectors,
to get the second Gateaux derivative of the l.h.s. This is known to be the {\em BKM} metric
\[\langle X,Y\rangle_{_B}=
{\rm Tr}\,\left(d\ell_+(\rho)d\ell_-(\rho)\right).\] This shows
that the duality between ${\cal X}$ and ${\cal X}^*$, given by the
trace form, can be expressed in terms of the {\em BKM} metric.

Given a Young function $\Phi$ defined on ${\cal X}$, we define the dual Young function
$\Phi^*$ on the dual space ${\cal X}^*$ by
\begin{equation}
\Phi^*(\rho_{_Y}):=\sup_{X\in{\cal X}}\left\{\langle
X,Y\rangle_{_B}-\Phi(X)\right\}, \hspace{.5in} \rho_{_Y}-\rho_{_0} \in{\cal X}^*.
\end{equation}
\begin{theorem}
{\em $\Phi^*$ is a Young function, it is lower semi-continuous in
the {\em BKM} metric, and Young's inequality
\begin{equation}
\Phi(X)+\Phi^*(\rho_{_Y})\geq\langle X,Y\rangle_{_B} \label{qyoung}
\end{equation}
holds for all $X,Y$.}
\end{theorem}
PROOF Clearly, $\Phi^*$ is even and vanishes at $Y=0$. For
convexity, let $\rho_1$ denote $\rho_{Y_1}$ etc., so that $\rho_1-\rho_{_0}$ is the
cotangent vector $d\ell_-(\rho_1)$. Then
\begin{eqnarray*}
\Phi^*(\lambda\rho_1+(1-\lambda)\rho_2)&=&\sup_X\left\{\lambda{\rm Tr}\,
\left(Xd\ell_-(\rho_1)\right)+(1-\lambda){\rm Tr}\,\left(Xd\ell_-(\rho_2)\right)-\Phi(X)\right\}\\
&\leq&\lambda\sup_X\left\{\langle
X,Y_1\rangle_{_B}-\Phi(X)\right\}+(1-\lambda)\sup_X\left\{
\langle X,Y_2\rangle_{_B}-\Phi(X)\right\}\\
&=&\lambda\Phi^*(\rho_1)+(1-\lambda)\Phi^*(\rho_2).
\end{eqnarray*}
It follows from $\Phi^*(\rho_{_0})=0$ and convexity that $\Phi^*(\rho_{_Y})\geq 0$
or is $\infty$.

$\Phi^*$, being the supremum of a family of continuous function
(indeed, continuous linear functions) is lower semi-continuous.
For Young's inequality, $\Phi^*(\rho_{_Y})$ being the supremum of $\langle
X,Y\rangle_{_B}-\Phi(X)$, cannot be smaller than any example. \hspace{\fill}{\em QED}

The double dual obeys $\Phi^{**}\leq \Phi$; for,
\begin{eqnarray*}
\Phi^{**}(X)&=&\sup_Y\left\{\langle X,Y\rangle_{_B}-\Phi^*(\rho_{_Y})\right\}\\
&=&\sup_Y\left\{\langle
X,Y\rangle_{_B}-\sup_{X^\prime}\left\{\langle
X^\prime,Y\rangle_{_B}-\Phi(X^\prime)\right\}\right\}\\
&=&\sup_Y\inf_{X^\prime}\left\{\langle
X-X^\prime,Y\rangle_{_B}+\Phi(X^\prime)\right\}\\
&\leq&\sup_Y\left\{\langle
X-X^\prime,Y\rangle_{_B}+\Phi(X^\prime)\right\}
\end{eqnarray*}
for all $X^\prime$. Choosing $X^\prime=X$ gives the inequality.

It follows that $\left(\Phi^*\right)^{**}\leq\Phi^*$. But we also
have the inequality the other way round:
\begin{eqnarray*}
\left(\Phi^*\right)^{**}(\rho_{_Y})&=&\left(\Phi^{**}\right)^*(\rho_{_Y})=\sup_X\left\{\langle
X,Y\rangle_{_B}-\Phi^{**}(X)\right\}\\
&\geq&\sup_X\left\{\langle X,Y\rangle_{_B}-\Phi(X)\right\}\\
&=&\Phi^*(\rho_{Y}),
\end{eqnarray*}
so $\left(\Phi^*\right)^{**}=\Phi^*$. This duality occurs because
$\Phi^*$ is lower semi-continuous. Indeed, $\Phi^{**}$ is the
lower semi-continuous version of $\Phi$ \cite {Ekeland}. From now
on, we shall assume that $\Phi$ is lower semi-continuous, so that
$\Phi^{**}=\Phi$.

We now consider the quantum analogue of the inequality
(\ref{holderorlicz}): the classical Young function
$\lambda\rightarrow\Phi(\lambda X)$ is continuous and increasing where finite. It
follows that the infimum in theorem~(\ref{qnorm}) is achieved at
$r=\|X\|_{_L}^{-1}$. Similarly for the dual Luxemburg norm. Now let $\|X\|_{_L}=1$ and $\|\rho_{_Y}-\rho_{_0}\|_{_{L^*}}=1$.
Then $\Phi(X)=1$ and $\Phi^*(\rho_{_Y})=1$, and by Young's inequality
(\ref{qyoung}),
\[
2\|X\|_{_L}\|\rho_{_X}-\rho_{_0}\|_{_{L^*}}=2=\Phi(X)+\Phi^*(\rho_{_Y})\geq
{\rm Tr}\,\left[X\left(\rho_{_Y}-\rho_{_0}\right)\right]\]
which multiplies up to give for tangent and cotangent vectors
\begin{equation}
\langle X,Y\rangle_{_B}\leq 2\|X\|_{_L}\,\|\rho_{_Y}-\rho_{_0}\|_{_{L^*}}.
\label{quorliczholder}
\end{equation}

\section{Conclusion}
We have argued that the information manifold in quantum theory
should consist of density operators $\rho$, some fractional power
of which is still of trace class. The topology on the manifold
should not be given by the trace norm. Instead, a 'hood of a given
state $\rho_{_0}$ should be given by $\epsilon$-bounded quadratic
forms; these were shown \cite{MRGRFS1} to make up a possible
analogue of the Cram\'{e}r class of random variables, in that
their Kubo-Mori expansion is analytic. This set of states carries
the $(+1)$ affine structure of Amari. A possible Young function,
related to the free energy, was introduced. The dual Young
function was shown to be finite on a set, the union of all
$p$-nearby states, and this carries the $(-1)$-affine structure of
Amari. The beginnings of Young theory (the {\em BKM}-metric, the
Luxemburg norms, Young's inequality and the H\"{o}lder-Orlicz
inequality) were derived.

Let us now complete ${\cal X}$ in the Luxemburg norm, and ${\cal
X}^*$ in the dual Luxemburg norm. The quantum H\"{o}lder-Orlicz
inequality (\ref{quorliczholder}) then shows that the bilinear
form between the spaces remains finite; we can therefore extend
the definition of the {\em BKM}-metric to the completions. The two
Banach spaces thus obtained contain only normal states. The
tangent and cotangent spaces are then complete and dual relative
to the Hilbert-Schmidt scalar product, and are furnished with the
$(\pm 1)$-affine structures. The tangent space then contains the
set of operators with finite Araki norm, and the cotangent space
contains the states which are perturbations of $\rho_{_0}$ by such
operators.

The Luxemburg norm becomes large when we add a perturbation such
that one of $e^{-H_{0}\pm X}$ nearly ceases to be of trace class.
In this way, the manifold consists of points that are in the
interior of some one-parameter exponential model. All states in
the manifold have finite entropy, and states near $\rho_{_0}$ have
finite relative entropy to $\rho_{_0}$.

One important property of the theory remains unproved: the
equivalence of the Luxemburg norms based on points $\rho_{_0}$ and
$\rho_{_X}$ for perturbations $Y$ lying in the overlaps of any 'hoods
of $\rho_{_0}$ with any 'hood of $\rho_{_X}$. It would also be nice for the
dual affine structures to be defined on the same space. In the classical case,
this was resolved by Grasselli \cite{MRG} in the subtheory obtained by
completing the space of bounded perturbations in the Luxemburg norm,
to obtain the (separable) Banach space $M$. Then the information manifold becomes
a Banach manifold modelled on $M$. In the quantum case, the
analogue of this space seems to be the completion in the Luxemburg norm
of the linear space consisting of perturbations of finite Araki norm.
One can ask whether this completion consists of only tiny forms.

\end{document}